\documentclass[letter,scriptaddress,twocolumn, pre,showpacs]{revtex4-1}
\usepackage{amssymb} 
	\usepackage{amsfonts}
	\usepackage[ansinew]{inputenc}
	\usepackage{epsfig}
 \usepackage{graphicx}
\begin{document}
%\begin{frontmatter}
\title{Stochastic modelling of dynamical systems with several attractors}
\author{M. Morillo and J. M. Casado}
%\email{casado@us.es}
\address{Area de F\'{\i}sica Te\'orica. Universidad de Sevilla\\
Apartado Correos 1085, 41080 Sevilla (Spain)}

\begin{abstract}
The usual Langevin approach to describe systems driven by noise fails to describe the long time behavior of systems with multiple attractors. The solution of the associated linear Fokker-Planck equation is always unique, even though it might show one or more maxima. In this context, it is customary to call transitions to changes in the shape of the equilibrium distribution function and relate the maxima to the attractors. Some years ago, a theory was developed for a system with interacting elements or subunits that, starting from the Langevin description of all the variables, leads to bifurcating \textquotedblleft one-particle\textquotedblright\, distribution functions when the number of elements tends to infinity.
In this paper, a mean-field hypothesis has been used to deal with systems with a finite number of elements. We carry out numerical simulations yielding bifurcation solutions for the probability density of a collective 
variable. We also compare the results of the mean-field hypothesis with those obtained with the Langevin approach for finite systems.    
\end{abstract}
\maketitle
\date{\today}
%\begin{keyword} 
%Stochastic models\sep Mean field approximation \sep Collective variable.
%\end{keyword}
%\end{frontmatter}
%\maketitle
\section{Introduction}
The addition of noise to nonlinear dynamical systems has unveiled a number of new phenomena with both fundamental and practical interest in many scientific areas \cite{CITAS1}. For deterministic systems with a single attractor, one can turn noise-free variables into fluctuating ones by introducing a \textquotedblleft fluctuacting force\textquotedblright\, to the equations of the model. This approach was pioneered by Langevin many years ago in his study of Brownian motion and, since then, it has been applied successfully to a great variety of problems in physics, chemistry and biology \cite{ARNOLD1981}. 

Nevertheless, the aforementioned program fails when applied to deterministic systems having a number of coexisting attractors, as it might be the case when the dynamics of the system is nonlinear. If we introduce the thermal effects (noise effects) by means of the usual Langevin construction, the corresponding Fokker-Planck equation for the probability density turns out to be a linear equation. Then, the possibility of having several distribution functions does not exist \cite{GANG1990}. It is possible that, as the value of the noise term is varied, the shape of the distribution function changes, so that for some range of noise values it can be multimodal, while it becomes monomodal for other noise values. When the distribution is multimodal one can think of associating each of its maxima to the corresponding deterministic attractors. But, we should keep in mind that, for a finite system, the multimodality of the distribution function indicates that the noise is able to connect the maxima, and consequently the attractors. This is in sharp contrast with the fact that the deterministic attractors are not connected. 

The behavior of the average value provided by the Langevin description can not even qualitatively coincide with that of the deterministic variables for all values of the parameters. For example, if the system symmetry leads to a multimodal stationary distribution with zero average value, the random trajectories will not remain indefinitely around one of the attractors, even for very small noise strengths. Instead, they will explore the entire space, yielding a zero average value. Furthermore, even for very small noise intensities, the fluctuations about the average value will be very large, in consonance with the fact that the noise forces the random trajectories to explore all the attractors. On the other hand, as the noise strength is increased, the random trajectories will explore more easily the entire space, thus connecting the attractors much more frequently and leading, for a sufficiently large noise value, to a monomodal stationary distribution. Consequently the fluctuations will tend to decrease in intensity as the noise strength is increased. This is bizarre as one is normally used to the idea that thermal fluctuations tend to increase as the temperature increases.

Consider a finite dynamical system characterized by, say, the non random variables $x_i,\, i=1,\ldots, N$ satisfying a set of nonlinear equations $\dot {x}_i=F_i(x_1,\ldots,x_j; \lambda) $ where $\lambda $ refers to some parameters. Imagine that the observation of the system leads to the conclusion that for some parameter values the system ---as characterized by a global variable $s(t)$--- has a single attractor, while for other set of parameter values, several attractors might be possible. Each of those attractors has its own basin of attraction. Then, the behavior of the system with time and its asymptotic state depends on its initial preparation. 

Imagine now that temperature plays a role, so that one observes that the transition between a single attractor to several attractors might depend not only on the system parameters, but also on the temperature. The presence of temperature will imply that the variables $x_i$ will no longer be deterministic, but they will have fluctuations. Let's denote by $X_i(t)$ the corresponding random variables. The question that arises is: what is the correct set of stochastic evolution equations compatible with the observations?

An example that comes to mind is the following: consider a permanent magnet which has been magnetized with its magnetic moment along a certain axis. If one does not apply any other external field, the magnetization will remain along the initial axis for as long as we observe it. Had we magnetized the sample along a different direction, it would stay like that for later times. Spontaneous transition of the magnetization among those directions are not observed as the temperature remains fixed. Now, suppose that we heat up the sample and the magnetization dissapears after a certain temperaure has been reached. Furthermore, assume that there are small fluctuations of the system magnetizations around each attractor value. Those fluctuations are not large enough to connect the two attractors during any reasonable amount of time. The nonstochastic dynamics can not describe these fluctuations. If the stochastic dynamics is modelled by the addition of a unbounded noise term representing the thermal effects, then we have fluctuations. But, if the Fokker-Planck description of the probability distribution is a linear one, the stationary distribution is unique. Thus, there must be fluctuations as large as one needs to connect the two attractors and the average magnetization will, sooner or later, be zero. If this is consistent with the experiments, fine. But if the system at the given temperature does not lose its magnetization in a spontaneous way, the Langevin (or linear Fokker-Planck) description is not correct.

\section{The model}
Let us consider a model system consisting of $N$ \textquotedblleft coordinates\textquotedblright\, $x_{i}$, whose dynamics are given by the set of coupled differential equations (in dimensionless form)
\begin{equation}
\dot{x}_{i}(t)=x_{i}(t)-x_{i}^{3}(t)+\frac{\theta}{N}\sum_{j=1}^{N}(x_{j}(t)-x_{i}(t)),
\label{EQ001}
\end{equation}
where $\theta$ is the strength of the coupling. We can rewrite this as
\begin{equation}
\dot{x}_{i}(t)=(1-\theta)x_{i}(t)-x_{i}^{3}(t)+\theta s(t),\qquad (i=1,2,\ldots, N).
\label{EQ0089}
\end{equation}
where $s(t)$ is a collective variable defined by
\begin{equation}
s(t)=\frac{1}{N}\sum_{j=1}^{N}x_{j}(t).
\label{EQ007}
\end{equation}

The numerical solution of Eqs.(\ref{EQ0089}) leads to an $s(t)$ that always ends in one of the attractors of the individual dynamics ($1$ or $-1$). However, the temporal evolution of the individual variables is very different to that of the $s(t)$ and depends on initial conditions. The collective variable is representative of the global behavior of the system even for the deterministic dynamics.

A stochastic version of the model was introduced some years ago by Kometani and Shimizu \cite{KOMETANI1975} in a biophysical context and it was later analyzed by Desai and Zwanzig\cite{DESAI1978}, Dawson\cite{DAWSON1983} and Shiino\cite{SHIINO1987} from a more general statistical mechanical perspective. The Langevin description amounts to add a stochastic term to each one of the deterministic equation, thus 
giving
\begin{equation}
\dot{X}_{i}(t)=(1-\theta)X_{i}(t)-X_{i}^{3}(t)+\theta S(t)+\xi_{i}(t),
\label{EQ0077}
\end{equation}
where 
\begin{equation}
S(t)=\frac{1}{N}\sum_{j=1}^{N}X_{j}(t),
\label{EQ0071}
\end{equation}
is the collective stochastic variable and $\xi_{i}(t)$ are Gaussian white noises with
\begin{equation}
\langle\xi_{i}(t)\rangle=0;\qquad \langle\xi_{i}(t)\xi_{j}(s)\rangle=2D\delta_{ij}\delta(t-s).
\label{EQ002}
\end{equation}   

An alternative formulation can be casted in terms of the linear Fokker-Planck equation for the joint probability distribution $f_{N}(x_{1},x_{2},\ldots,x_{N},t)$: 
\begin{eqnarray}
\frac{\partial f_{N}}{\partial t}&=&\sum_{i=1}^{N}\frac{\partial }{\partial x_{i}}\Big(\frac{\partial U}{\partial x_{i}}f_{N}\Big)\nonumber\\
&+& D\sum_{i=1}^{N}\frac{\partial^{2}f_{N}}{\partial x^{2}_{i}},\end{eqnarray}
where $U$ is the potential energy relief,
\begin{equation}
U=\sum_{i=1}^{N}V(x_{i})+\frac{\theta}{4N}\sum_{i=1}^{N}\sum_{j=1}^{N}(x_{i}-x_{j})^{2},
\label{relief}
\end{equation}
with the single particle potential
\begin{equation}
V(x)=-\frac{x^{2}}{2}+\frac{x^{4}}{4}.
\label{singlepotential}
\end{equation}

A problem with the above stochastic description is that, for any finite system, the joint probability
distribution satisfies a linear Fokker-Planck equation. Consequently, the multiplicity of solutions in the deterministic limit does not manifest in the stochastic description as the joint probability distribution is unique. For some values of the parameters, the unique stationary distribution for the collective variable might show two peaks, reminiscent of the two deterministic attractors. Regardless of the initial conditions, the stochastic trajectories will explore the two attractors, at variance with the deterministic limit where the two attractors are independent. 

As pointed out by Desai and Zwanzig, one can propose a cumulant moment expansion treatment of the
linear Fokker-Planck equation. Truncation of the cumulant expansion at some reasonable level (say, the Gaussian level) leads to a nonlinear evolution equation for the averages. For some values of the parameters, this truncation provides an adequate qualitative description ($\theta > 1) $ for the model at hand, but for other values of the parameter, any truncation leads to wrong qualitative results \cite{Drozmor1996}. 

\begin{figure}
\centerline{\epsfig{figure=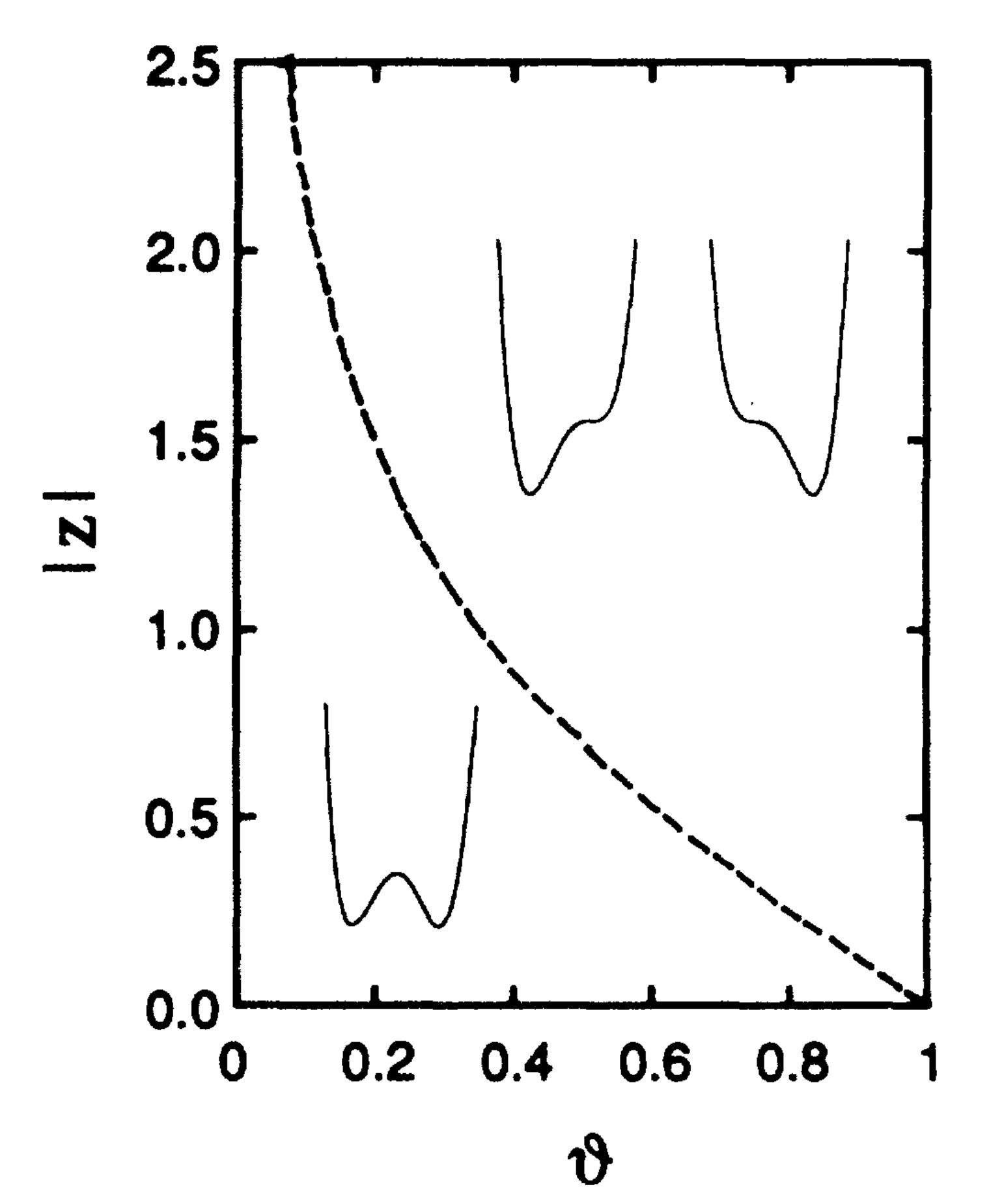,height=6cm}}
\caption{\label{FIG001}\small Bifurcation line for the DZ model with $0<\theta<1$. The variable $|z|=|\theta-1|/\sqrt{2D}$ has been used instead of $D$. Under the line there is a unique, symmetric, one-particle equilibrium distribution function that is bistable. Above it there are two equilibrium probability distributions with nonzero mean values.}
\end{figure}

In the limit $N\to\infty$, Desai and Zwanzig \cite{DESAI1978} were able to write a Nonlinear Fokker-Planck Equation (NLFPE) (see also \cite{FRANK,KURSTEN2016}) describing the behavior of the \textquotedblleft one-particle\textquotedblright\, probability distribution, 
\begin{eqnarray}
\label{nlfp1}
\frac{\partial f_{1}(x,t)}{\partial t}&=&\frac{\partial}{\partial x}\Big\{\big[(\theta-1)x+x^{3}-\theta\langle x(t)\rangle\big]f_{1}(x,t)\Big\}\nonumber\\&+&D\frac{\partial^{2}f_{1}(x,t)}{\partial x^{2}},
\end{eqnarray}
where
\begin{equation}
\langle x(t)\rangle=\int_{-\infty}^{\infty}xf_{1}(x,t)dx.
\end{equation}
In this approach, the infinite system undergoes an order-disorder phase transition for some values of parameters $\theta$ and $D$. This phase transition is signaled by a stochastic bifurcation leading to multiple equilibrium distributions $f_{eq}(x)$. In Fig.\ref{FIG001} the bifurcation line separating the zones of one and three stationary solutions in the parameter space $\theta$ and $|z|=|\theta-1|/\sqrt{2D}$ has been depicted \cite{DESAI1978}. 

Let us introduce an alternative approach to the description of a system with a finite number of elements. We will substitute the stochastic variable $S(t)$ in the finite set of equations (\ref{EQ0077}) by its mean field version in the Weiss spirit, thus writing
 \begin{equation}
\dot{X}_{i}(t)=(1-\theta)X_{i}(t)-X_{i}^{3}(t)+\theta \langle S(t)\rangle+\xi_{i}(t),
\label{EQ100}
\end{equation}
where $\langle\ldots\rangle$ represents an average over the noise. The interactions between the $X_i(t)$ variables are then replaced by the average value of the collective one. Note that within this description, the joint probability distribution $f_{N}(x_{1},x_{2},\ldots,x_{N},t)$ factorizes as a product of single particle distributions, each of them satisfying a nonlinear Fokker-Planck, Eq.\ (\ref{nlfp1}).
Also note that $\langle S(t) \rangle = \langle x(t) \rangle $. The probability distribution for the collective variable is defined by
\begin{equation}
P(s,t) = \int dx_1\cdots dx_N\; \delta \Big (s-\frac 1N \sum x_i \Big) f_N(x_1,\cdots,x_N,t) 
\label{pst}
\end{equation}

A purpose of this work is to study the distribution function $P_{eq}(s)$ for the collective variable of a finite system 
in the range of parameters where there are multiple attractors and to compare the results with those obtained with the finite set of Langevin equations, Eq.\  (\ref{EQ002}). We are also interested in analyzing the relaxation of the average collective behavior towards a stationary situation, starting from different initial conditions. 

\section{Numerical results}
We have not been able to find a suitable analytical approximation for the process $S(t)$. Thus, we will rely on numerical simulations to get the results presented here. 
In what follows, we have chosen $\theta=0.5$, thus allowing the noise strength $D$ to be the only bifurcation parameter of the system. Other values of the parameter $\theta $ lead to qualitatively analogous results.

We have solved numerically the systems of equations, Eqs. (\ref{EQ0077}) and (\ref{EQ100}) for a small number of particles $(N=11)$ using different realizations of the noise. After a convenient relaxation time so that we are sure a stationary probability distribution has been reached, we record during a long time interval and for all the random trajectories, the times that the collective variable is within
the different bins of the suitably discretized collective variable.
In this form we have been able to construct histograms that approximate the equilibrium probability distributions $P_{eq}(s)$ for the Langevin and the mean-field descriptions. 

\begin{figure}[h]
\centerline{\epsfig{figure=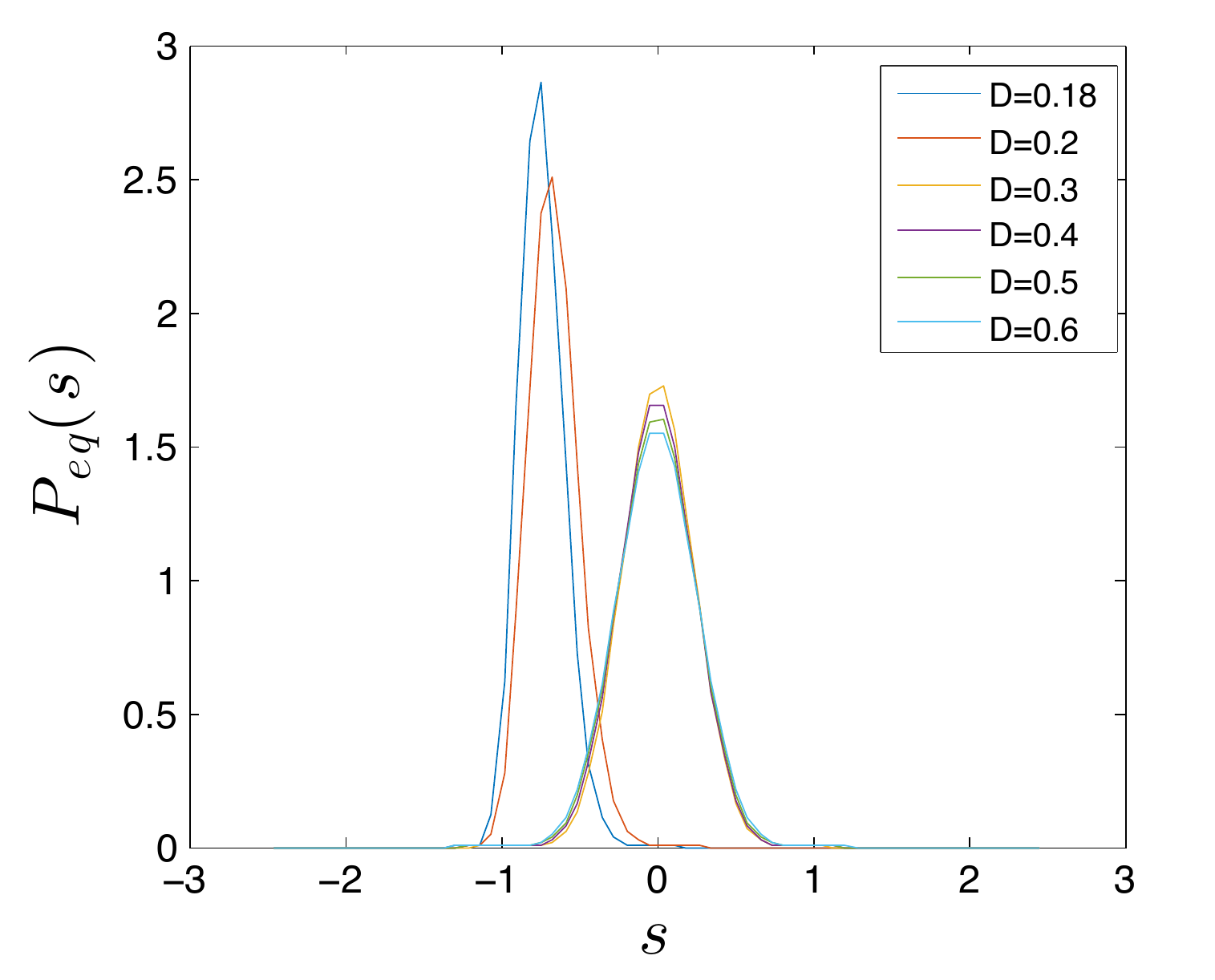,height=6cm}}
\caption{\label{PSMF}\small The equilibrium distribution for the collective variable in the mean-field description, Eq. (\ref{EQ100}), for a system with $N=11$ elements. The initial conditions of the individual coordinates $x_{j}(0); j=1,\ldots, N$ leads to $S(0) < 0$. }
\end{figure}

In Fig.\ref{PSMF} we have represented the equilibrium distribution $P_{eq}(s)$ for the collective variable in the mean-field description for some values of the noise strength $D$. In all cases the initial conditions of the individual elements were chosen so that $S(0)<0$. As we can observe, there is a range in $D$ where $P_{eq}(s)$ is a zero-centered, monomodal distribution while, as the value of $D$ decreases, the distribution peaks around a finite (negative) value of $s$. For initial conditions with $S(0)>0$, and the same small values of the noise strength, a symmetric $P_{eq}(s)$ (not depicted) is reached that peaks around a positive value of $s$. 

\begin{figure}
\centerline{\epsfig{figure=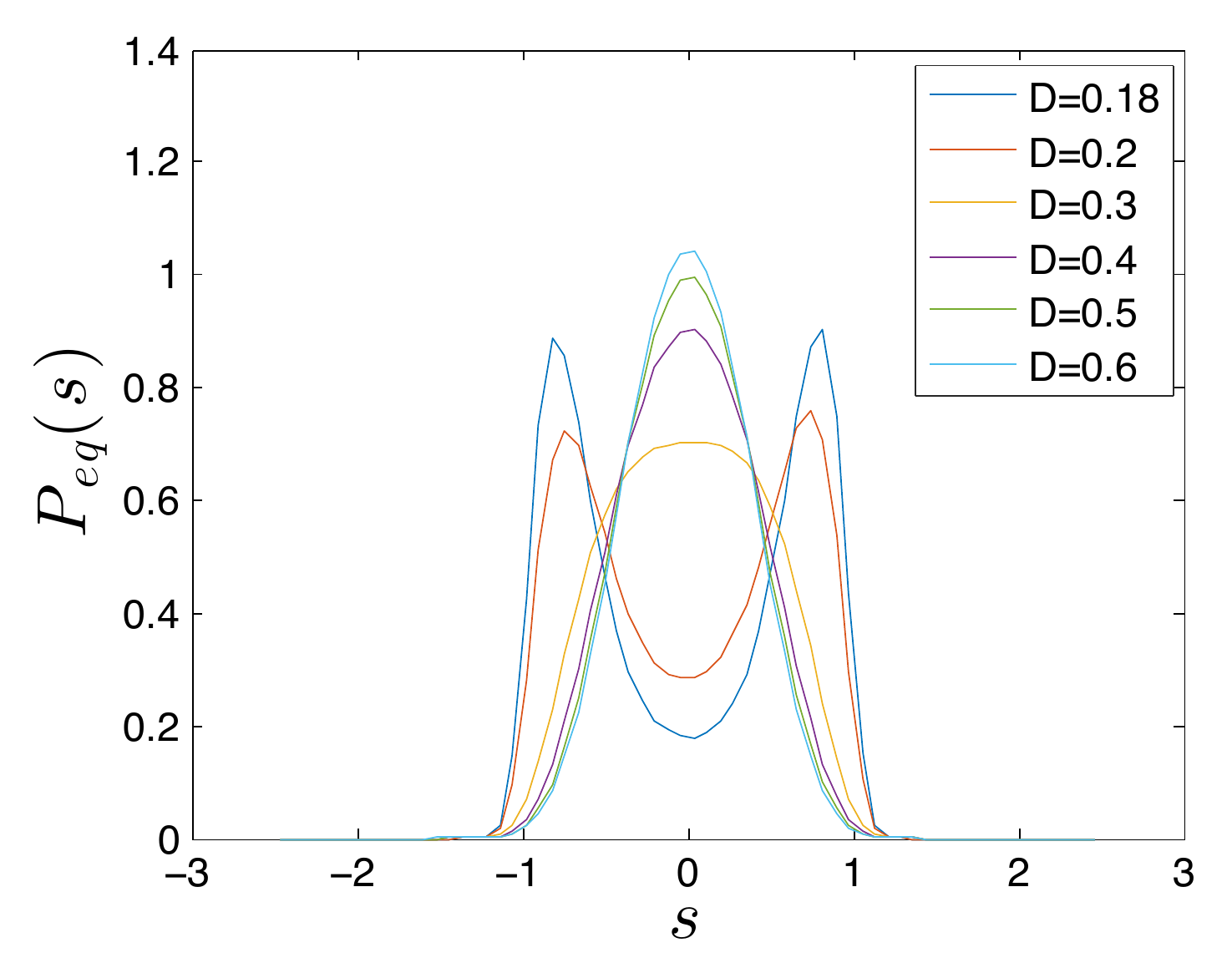,height=6cm}}
\caption{\label{PSSTO}\small The equilibrium distribution  for the collective variable in the Langevin description, Eq. (\ref{EQ0077}), of a system with $N=11$ elements. The initial condition for the collective variable is $S(0) < 0$.}
\end{figure}

The above behavior is in contrast with that obtained by using the Langevin description. In this case, as shown in Fig. \ref{PSSTO}, the equilibrium distribution $P_{eq}(s)$ is always unique regardless of whether the initial preparation yields $S(0)<0$ or $S(0)>0$. The equilibrium distribution is always zero centered, although its shape changes from a single peak distribution at high noise intensities to a bimodal one as the noise strength is decreased. This is to be expected as the strong noise blends the two symmetric deterministic attractors into a single equilibrium point with the highest probability. For low noise values, the random trajectories are essentially fluctuating around the deterministic attractors, except for some sporadic transitions between them. As the system size increases, the two maxima of the bimodal distribution become narrower while the local minima around $s=0$  becomes much smaller.
It is not surprising, then, that in the limit $N\to\infty$, the two peaks are completely separated and one finds two equilibrium monomodal distributions, either one of them being reached depending on the initial conditions.   

\begin{figure}
\centerline{\epsfig{figure=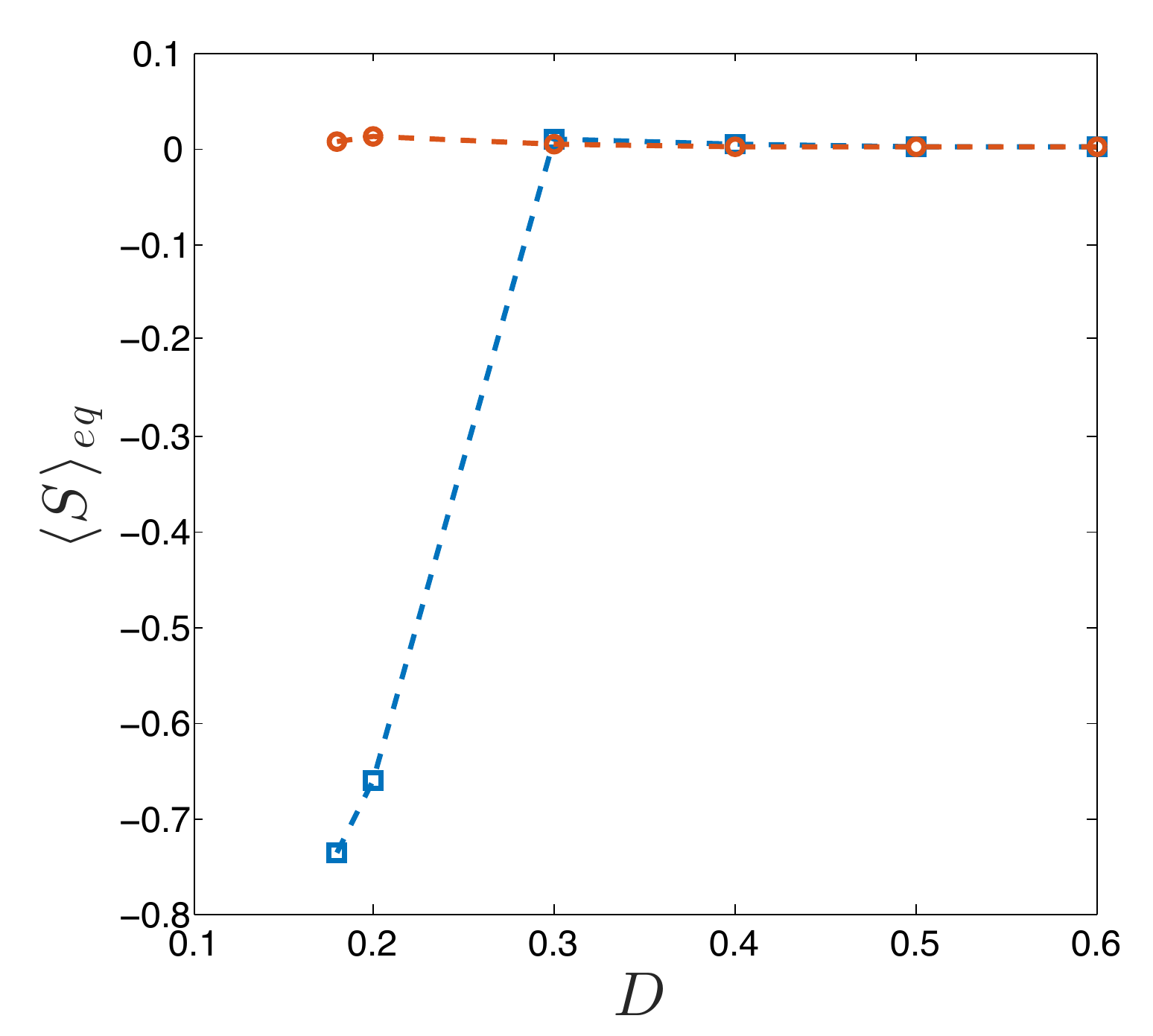,height=6cm}}
\caption{\label{AVG}\small The equilibrium average value of the collective variable in the mean-field description (blue) and within the Langevin description (red). The initial condition is $S(0) < 0$. For $S(0)>0$ a symmetrical curve would have been obtained for the mean-field case (no depicted here).}
\end{figure}

A further comparison has been made between these two descriptions by calculating the dependence of the average value $\langle S\rangle_{eq}$ and dispersion $\sigma_{s}=\sqrt{\langle S^{2}\rangle_{eq}-\langle S\rangle_{eq}^2}$ on $D$ for both cases. We have numerically evaluated the integrals leading to  $\langle S\rangle_{eq}$ and $\langle S^{2}\rangle_{eq}$ using the histograms.
As we can observe in Fig. \ref{AVG}, the average value is always zero for the Langevin description whether or not the distribution is monomodal o bimodal, due tho the symmetric character of the distribution. For the mean-field description, however, the average value bifurcates from zero to finite (positive or negative) values of $s$ as $D$ is decreased.  

\begin{figure}
\centerline{\epsfig{figure=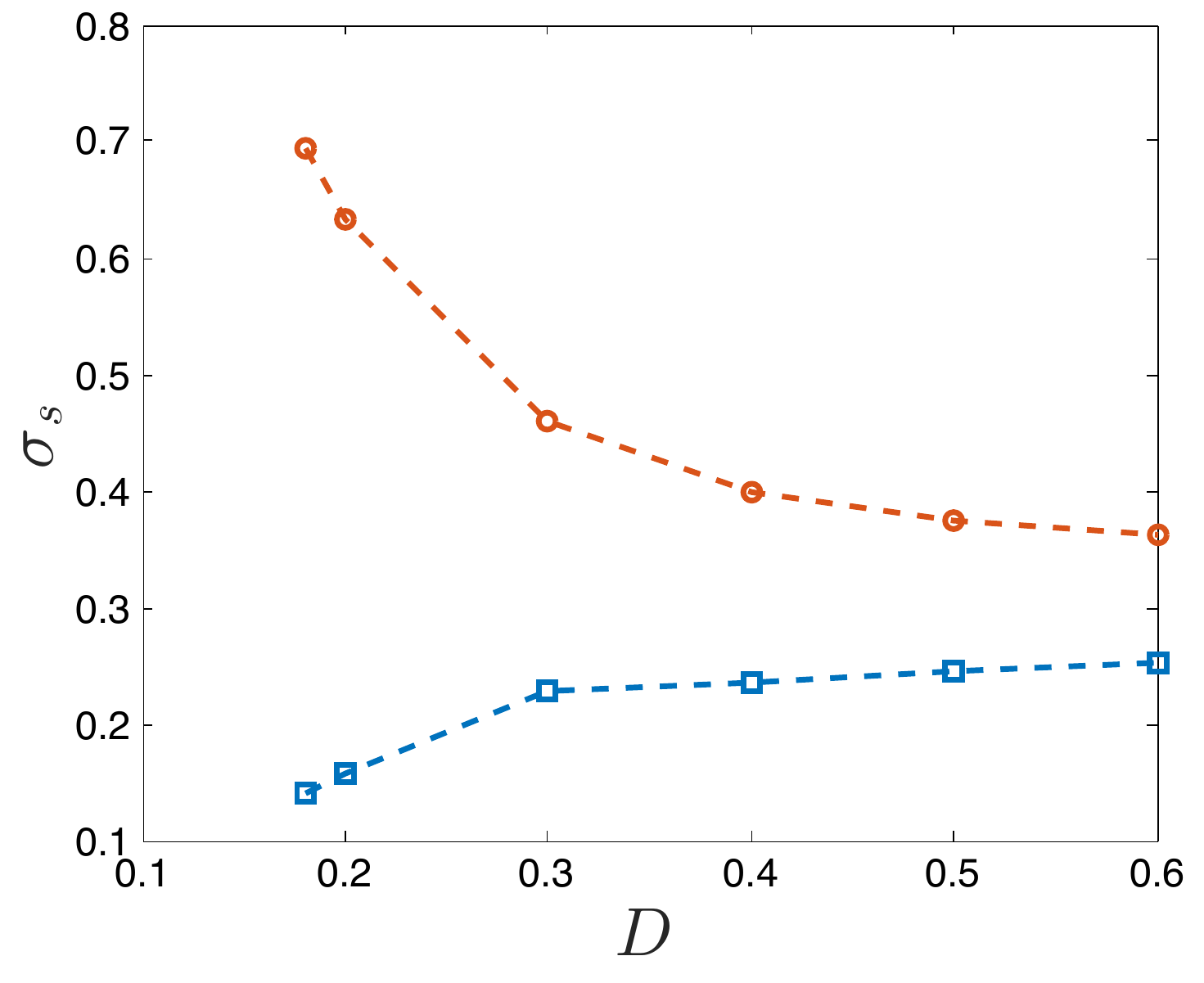,height=6cm}}
\caption{\label{CUM}\small The dispersion of the collective variable in the mean-field description (blue) and in the Langevin description (red). The initial condition is $S(0) < 0$.}
\end{figure}

The behavior of the dispersion as $D$ changes is also very different in both descriptions. As we can see in Fig.\ref{CUM}, as the noise strength is increased, the value of $\sigma_{s}$ increases in the case of mean-field dynamics and decreases in the Langevin description. The fact that the dispersion decreases as the noise increases in the Langevin description is tied to the fact that the corresponding equilibrium distribution $P_{eq}(s)$ changes from having a single peak for large noise values to be bimodal at low noise intensities.

\begin{figure}
\centerline{\epsfig{figure=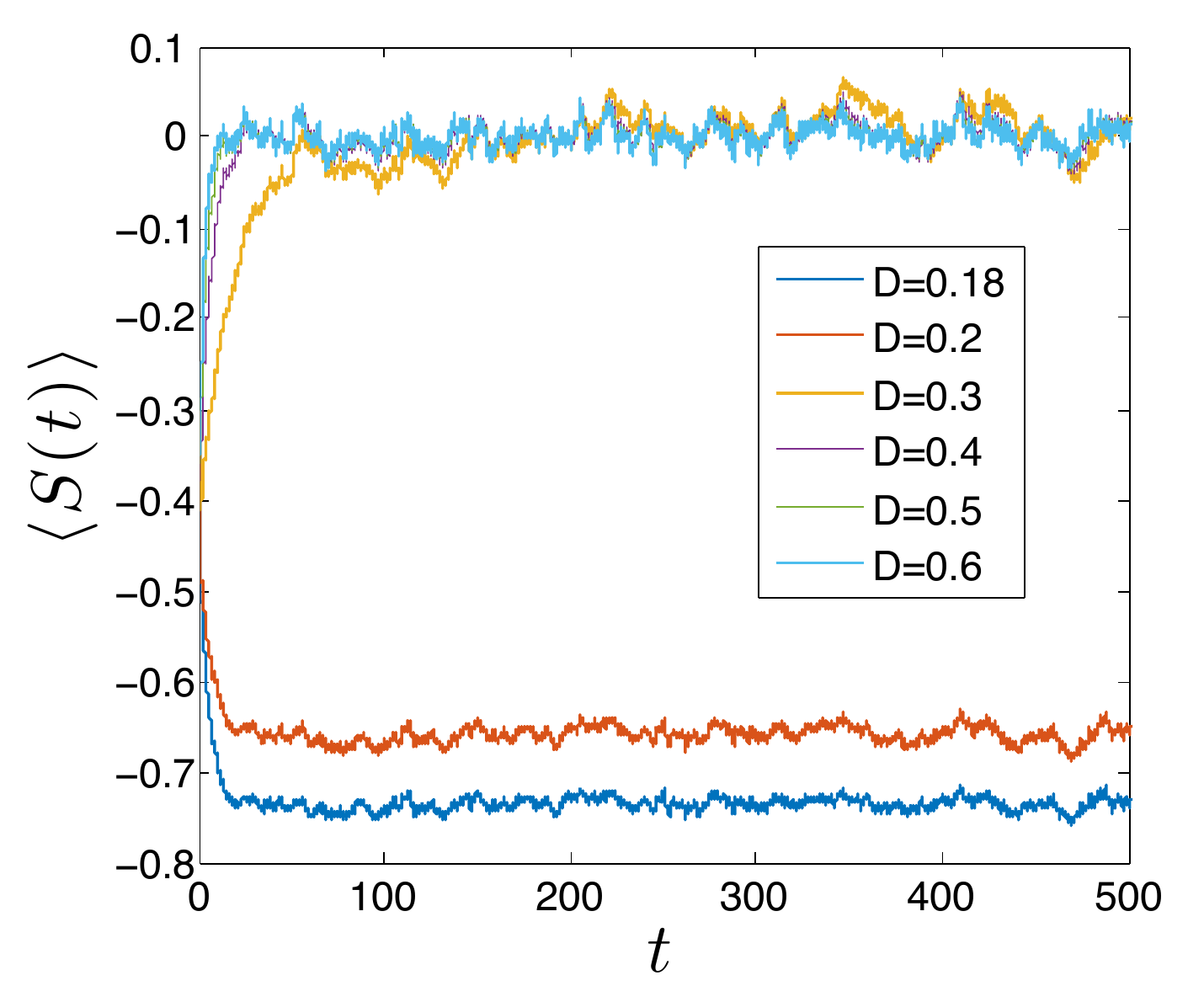,height=6cm}}
\caption{\label{REL1}\small Relaxation towards equilibrium of $\langle S(t)\rangle$ in the mean-field description for several values of noise strength. The initial condition corresponds to $S(0) < 0$.}
\end{figure}

\begin{figure}
\centerline{\epsfig{figure=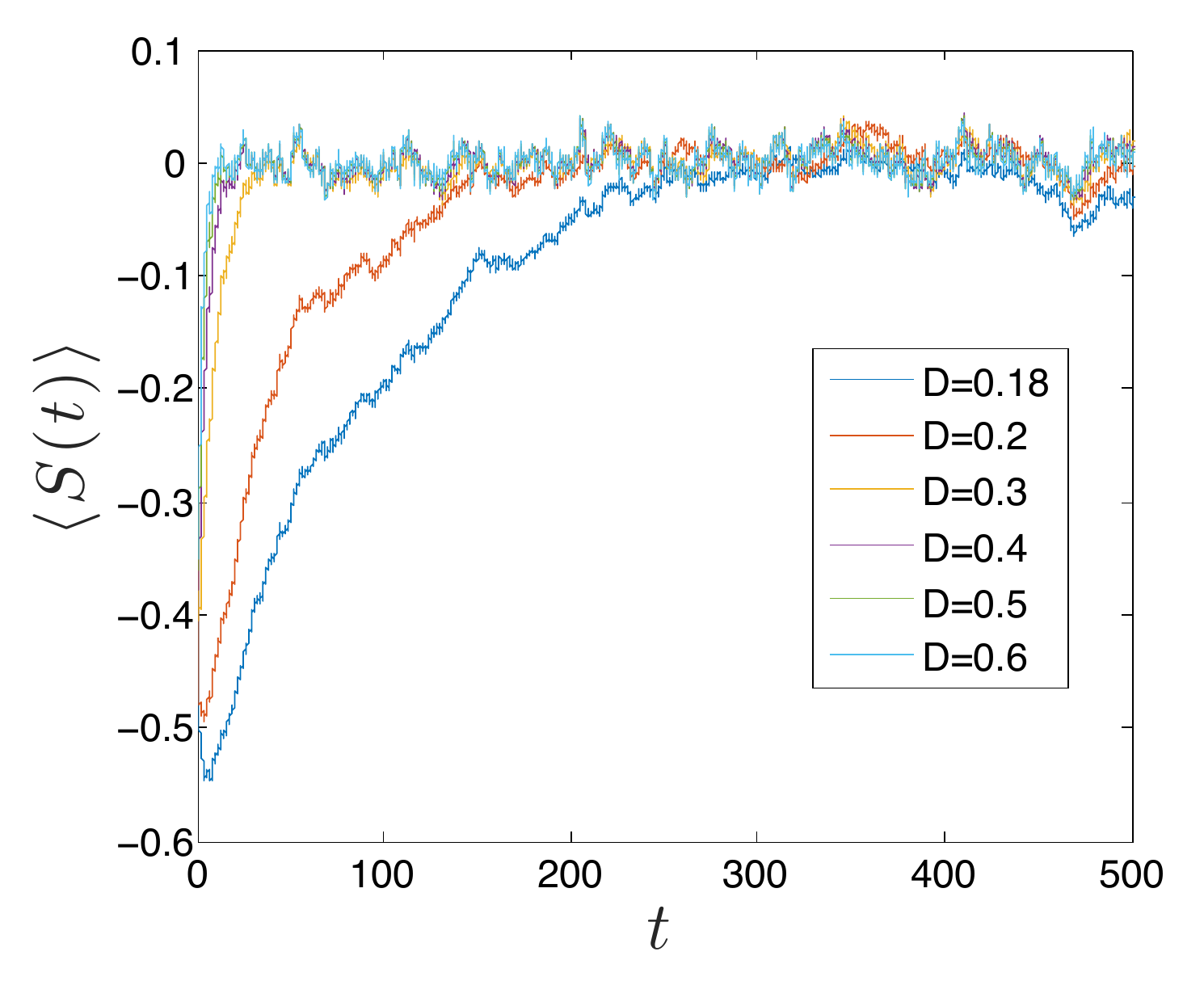,height=6cm}}
\caption{\label{REL2}\small Relaxation towards equilibrium of $\langle S(t)\rangle$ in the Langevin description for the same initial conditions and parameter values.}
\end{figure}

Another aspect of our analysis is that of relaxation towards equilibrium. It is useful to compare the relaxation time of the average behavior with that of the deterministic relaxation and also to compare the relaxation times in both stochastic descriptions for the same values of the noise strength. This last comparison can be made by observing Figs. \ref{REL1} and \ref{REL2}. For small noise values, the differences in the relaxation times are noteworthy.  

\begin{figure}
\centerline{\epsfig{figure=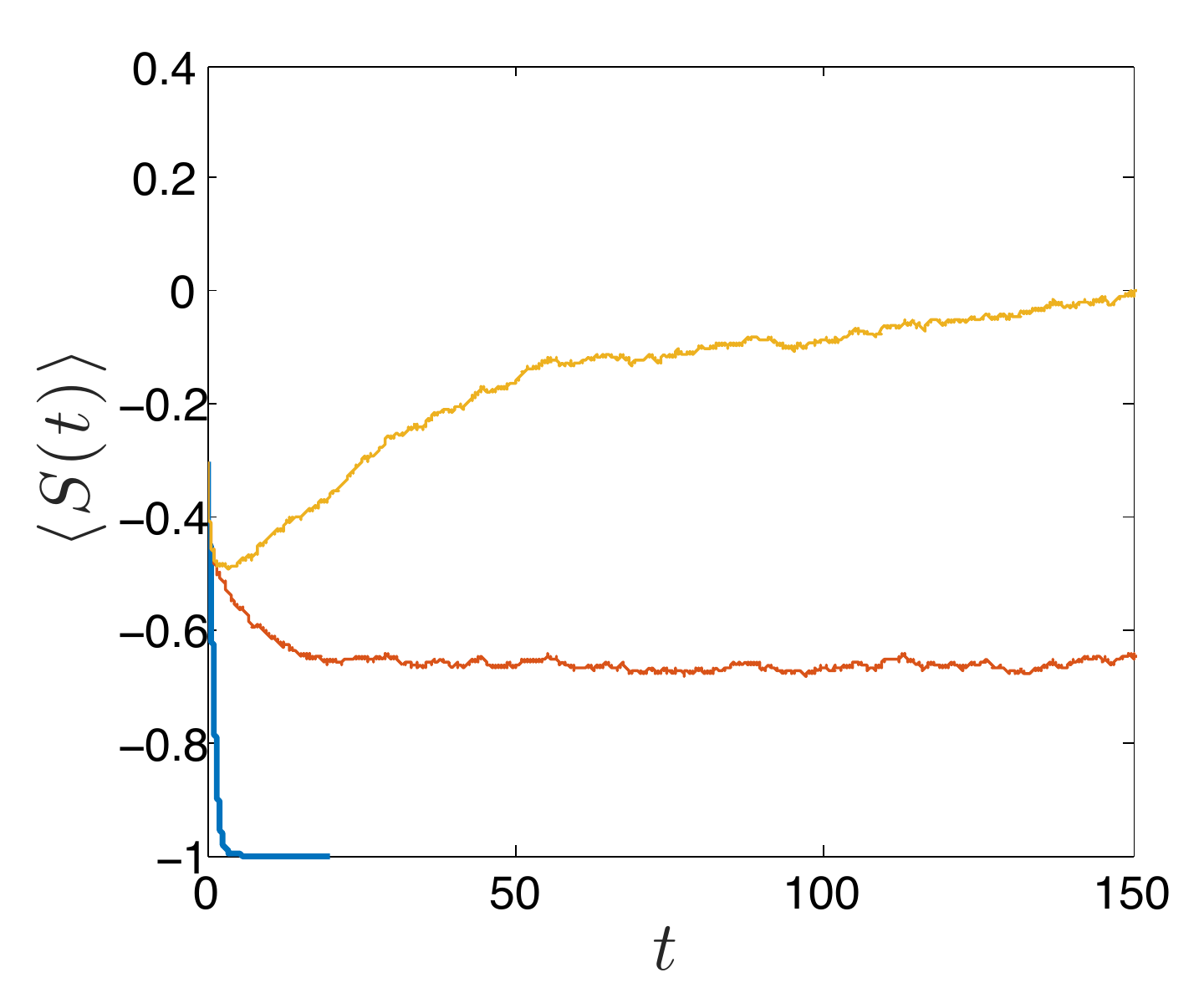,height=6cm}}
\caption{\label{REL3}\small Relaxation towards equilibrium of $\langle S(t)\rangle$ in the deterministic limit $D=0$ (blue line), in the Langevin description (yellow line) $D=0.2$, and the mean-field description (red line) $D=0.2$, starting from  the same initial conditions.}
\end{figure}

In Fig. \ref{REL3} we depict an example of the relaxation of the average collective variable as given by the two stochastic descriptions analyzed above and by the deterministic dynamics. As we can see, the Langevin dynamics shows an average behavior towards a zero value, even though the deterministic dynamics shows a relaxation towards $s=-1$, the attractor reached by the initial condition $S(0)=-0.3$. The mean-field dynamics, on the other hand, indicates a relaxation towards a negative value, which differs from the deterministic attractor due to the renormalization of the deterministic nonlinear dynamics by the coupling of noise and nonlinearity, a well known effect in stochastic nonlinear systems \cite{NORDHOLM1974}.

\section{Conclusions}
We have discussed in this paper a possible way to analyze noise effects in the description of finite size nonlinear systems which, in the deterministic limit,
would present two disconnected stationary attractors. The variable of interest is a collective variable characterizing the whole system. Due to the nonlinear nature of the problem, an analytical solution is out of the question and we have relied on  numerical simulations.

We have compared the characteristics of the equilibrium distribution function of the collective variable for two different approaches to describe the fluctuating dynamics of a finite system with interacting subunits. A possible description is to add a noise term to each of the individual deterministic degrees of freedom as it is typically done in a 
Langevin description. As an alternative, we propose a stochastic description leading to nonlinear Fokker-Planck equations. In this approach, we make use of an extension of the Weiss procedure to explain the paramagnetic-ferromagnetic transition, and incorporate the collective variable average value into the stochastic dynamics of each variable. 

Our numerical results show that the equilibrium probability distribution for the collective variable $P_{eq}(s)$ in the mean-field approach presents a bifurcating character signaling the occurrence of a order-disorder transition. For high values of the noise strength,  $P_{eq}(s)$ is an almost Gaussian distribution centered around zero, regardless of the initial condition. As the noise strength is decreased, two stable stationary distributions centered around positive or negative values  are possible. Which one of them is reached depends on the initial preparation.

The symmetrical mean values obtained from the bifurcating solutions are not identical to the stationary values in the deterministic description. This is not surprising, due to the renormalization effects associated to the noise, nonlinearity and cooperativity.
%\end{document}
%\newpage

\end{document}